\documentclass[preprint]{aastex}
\usepackage{emulateapj5}
\usepackage{epsf}
\usepackage{onecolfloat}
\begin{document}

\submitted{The Astrophysical Journal, submitted}
\vspace{1mm}
\slugcomment{{\em The Astrophysical Journal, submitted}} 

\twocolumn[

\title{The Radial distribution of galaxies in $\Lambda$CDM clusters}

\author{Daisuke Nagai\altaffilmark{1} and Andrey
V. Kravtsov\altaffilmark{1}} \affil{Department of Astronomy and
Astrophysics, Kavli Institute for Cosmological Physics,\\ 5640 South
Ellis Ave., The University of Chicago, Chicago, IL 60637}

\begin{abstract}  
  We study the radial distribution of subhalos and galaxies using
  high-resolution cosmological simulations of galaxy clusters formed
  in the concordance $\Lambda$CDM cosmology.  In agreement with
  previous studies, we find that the radial distribution of subhalos
  is significantly less concentrated than that of the dark matter,
  when subhalos are selected using their present-day gravitationally
  bound mass.  We show that the difference in the radial distribution
  is not a numerical artifact and is due to tidal stripping.  The
  subhalos in the cluster core lose more than $70\%$ of their initial
  mass since accretion, while the average tidal mass loss for halos
  near the virial radius is $\approx 30\%$. This introduces a radial
  bias in the spatial distribution of subhalos when the subhalos are
  selected using their tidally truncated mass. We demonstrate that the
  radial bias disappears almost entirely if subhalos are selected
  using their mass or circular velocity at the accretion epoch. The
  comparisons of the results of dissipationless simulations to the
  observed distribution of galaxies in clusters are therefore
  sensitive to the selection criteria used to select subhalo samples.
  Using the simulations that include cooling and starformation, we
  show that the radial distribution of subhalos is in reasonable
  agreement with the observed radial distribution of galaxies in
  clusters over the entire radial range probed by the simulations,
  $0.1<R/R_{\rm 200}<2.0$, if subhalos are selected using the stellar
  mass of galaxies they harbor. The radial bias is minimized in this
  case because the stars are located in the centers of dark matter
  subhalos and are tightly bound.  The stellar mass of an object is
  therefore approximately conserved as the dark matter is stripped
  from the outer regions. Nevertheless, the concentration of the
  radial distribution of galaxies is systematically lower than that of
  the dark matter.  Finally, we find that the baryon dissipation does
  enhance the survival of subhalos, especially in the cluster cores.
  However, the effect is relatively weak because the baryon
  dissipation affects the density distribution only at small radii.
\end{abstract}


]

\altaffiltext{1}{{\tt daisuke,andrey@oddjob.uchicago.edu}}

\section{Introduction}
\label{sec:intro}

Understanding the spatial distribution of galaxies and matter is one
of the main goals of observational and theoretical cosmology.  At
large scales, the galaxy and matter distributions on average are
thought to be simply related by a constant scale-independent bias,
with the constant depending on the galaxy mass
\citep[e.g.,][]{mo_white96,scherrer_weinberg98}. Clustering of
galaxies at small scales ($r\lesssim 3h^{-1}\ \rm Mpc$), on the other
hand, is thought to be determined by the abundance and radial
distribution of galaxies within dark matter halos \citep[e.g.,
see][for a recent review]{cooray_sheth02}.

While the abundance of galaxies in halos of different mass have
recently been extensively quantified in cosmological simulations
\citep[e.g.,][]{seljak00,guzik_seljak01,white_etal01,berlind_etal03,
kravtsov_etal04}, their radial distribution is less well understood.
Theoretical analyses often simply assume that galaxies trace the
radial profile of dark matter.  However, it is easy to imagine that
dynamical friction, tidal disruption, and morphological transformation
could modify the distribution of galaxies with respect to matter. In
addition, the radial distribution of galaxies should clearly be
sensitive to how galaxies are selected. For example, the distribution
of early type red galaxies in clusters is more centrally concentrated
than that of late type blue galaxies \citep[e.g.,][]{goto_etal04}.
New cluster samples based on large galaxy surveys
\citep{carlberg_etal97,vandermarel_etal00,lin_etal04,miller_etal04}
should shed light on these issues, but the interpretation of
observational results will require comparisons with theoretical
expectations.

Properties of galaxy-size dark matter halos in groups and clusters (or
{\it the subhalos}) have been the subject of many recent studies,
which used a new generation of high-resolution dissipationless
simulations not affected by the ``overmerging'' problem
\citep{ghigna_etal98,tormen_etal98,klypin_etal99,colin_etal99,okamoto_habe99,
colin_etal00,ghigna_etal00,springel_etal01,taffoni_etal03,delucia_etal04,tormen_etal04,kravtsov_etal04b,diemand_etal04,gao_etal04a,reed_etal04}.
One of the main results is that the radial distribution of subhalos is
less concentrated than that of dark matter
\citep{ghigna_etal98,colin_etal99,ghigna_etal00,springel_etal01,
delucia_etal04,gao_etal04b}. The subhalos also appear to have a
significantly more extended and shallower radial distribution compared
to the observed distribution of galaxies in clusters, a point recently
emphasized by \citet{diemand_etal04} and \citet{gao_etal04b}.

Although questions were raised about remaining overmerging in the
inner regions of halos \citep{taylor_etal03}, \cite{diemand_etal04}
used convergence tests to show that the subhalo distribution is not
affected by resolution but is due to real physical merging of subhalos
in dissipationless simulations (see also \S~\ref{sec:convtest} below).
They also argued that the physical overmerging in dissipationless
simulations results in a factor of two underestimate of the number of
galactic subhalos in cluster-size systems.

One of the obvious omissions in dark matter only simulations is the
lack of dissipation and starformation. After all, the observed samples
of galaxies are most often selected using their luminosity or an
observational proxy for stellar mass, not a total gravitationally
bound mass. In addition, when baryons cool and condense in the cores
of dark matter halos, they increase the inner density of their host,
which can make it more resistant to tidal disruption
\citep{white_rees78}.  It is therefore important to test whether
inclusion of cooling and starformation, the processes critical for
realistic modeling of galaxy formation, results in significantly
different galaxy populations in clusters.

Although a number of studies during the last decade used gasdynamics
cosmological simulations with cooling to study galaxy clustering on
small scales \citep[e.g.,][]{katz_etal92,pearce_etal99,blanton_etal00,
  yoshikawa_etal01,weinberg_etal04}, only a few analyses directly
addressed the radial distribution of galaxies in halos.
\cite{frenk_etal96} used simulations with cooling and starformation to
study distribution and dynamics of galaxies in clusters. They found
that stellar systems formed in the simulations were more centrally
concentrated than dark matter. This, however, was the case only in the
runs with starformation. When gas was allowed to cool without forming
stars, the simulations appeared to suffer from the ``overmerging''
problem. More recently, \cite{berlind_etal03} analyzed the radial
distribution of galaxies, identified as dense baryonic clumps, in
halos of different mass in SPH simulations. These authors found that
galaxies selected to have baryonic masses above a certain threshold
have a somewhat more extended distribution compared to the dark
matter.

A number of studies used a hybrid approach of combining
high-resolution $N$-body simulations with semi-analytic models of
galaxy formation to study galaxy populations in clusters and groups
\citep{springel_etal01,
  okamoto_nagashima01,kravtsov_etal04b,gao_etal04b}.  All of these
models imply that radial distribution of halos is more extended than
that of the luminous galaxies. By necessity, semi-analytic studies
make certain assumptions about dynamical evolution of the stellar
component, not included directly in simulations. For example,
\citet{springel_etal01} and \citet{gao_etal04b} find that for a large
fraction of objects in the cluster core the stellar systems may
survive as distinct galaxies, even after their DM halos are disrupted.
\citet{kravtsov_etal04b} used the tidal force measured in simulations
to estimate the amount of tidal heating and mass loss of galaxies
semi-analytically.

It is clearly important to address the radial distribution of subhalos
and galaxies in self-consistent high-resolution cosmological
simulations of cluster formation. In this paper, we present such a
study.  We compare results of cluster simulations with gasdynamics,
cooling, and starformation to the dissipationless simulations started
from the same initial conditions. We test the convergence of our
results by comparing results of two simulations of the same cluster
with an order of magnitude different resolution.

The paper is organized as follows. In the following two sections, we
describe the numerical simulations and halo finding algorithm used in
our analysis.  In \S~\ref{sec:convtest} we present the convergence
tests.  We present the results of dissipationless simulations in
\S~\ref{sec:Rad_Nbody} and gasdynamics simulations in
\S~\ref{sec:Rad_Hydro}.  We summarize our results and conclusions in
\S~\ref{sec:discussion}.

\section{Simulations}
\label{sec:sim}

In this study, we analyze high-resolution cosmological simulations of
eight group and cluster-size systems in the ``concordance'' flat
{$\Lambda$}CDM model: $\Omega_{\rm m}=1-\Omega_{\Lambda}=0.3$,
$\Omega_{\rm b}=0.043$, $h=0.7$ and $\sigma_8=0.9$, where the Hubble
constant is defined as $100h{\ \rm km\ s^{-1}\ Mpc^{-1}}$, and
$\sigma_8$ is the power spectrum normalization on $8h^{-1}$~Mpc scale.
The virial masses of cluster systems we consider range from $\approx
7\times10^{13}h^{-1}{\ \rm M_{\odot}}$ to $3\times 10^{14}h^{-1}{\ \rm
M_{\odot}}$. The simulations were done with the Adaptive Refinement
Tree (ART) $N$-body$+$gasdynamics code \citep*{kravtsov99,
kravtsov_etal02}, an Eulerian code, which uses the adaptive refinement
in space and time, and (non-adaptive) refinement in mass
\citep{klypin_etal01} to reach the high dynamic range required to
resolve cores of halos formed in self-consistent cosmological
simulations.

For each cluster, we analyze two sets of simulations started from the
same initial conditions but with different physical processes
included. The first set of simulations followed dynamics of dark
matter only. To study effects of resolution, one of the clusters was
re-simulated with eight times more particles and with higher spatial
resolution \citep{tasitsiomi_etal04a}.  We will be denoting lower and
higher resolution simulations of this cluster as LR and HR runs.  Both
LR and HR runs used a $256^3$ uniform root grid covering the
computational box of $80h^{-1}$~Mpc, but different numbers of
particles and refinements. The LR run has an effective mass resolution
of $m_{\rm p}^{\rm LR}=3.16\times 10^8h^{-1}M_{\odot}$, corresponding
to $512^3$ particles in the box, and reached 8 levels of refinement,
corresponding to the size of the highest refinement level cell of
$1.2h^{-1}$~kpc. The HR run has eight times more particles ($m_{\rm
p}^{\rm HR} = 3.95\times 10^7h^{-1}\ M_{\odot}$). In this simulation,
the smallest cell size reached was $0.6h^{-1}$ comoving kpc.

The gasdynamics simulations started from the same initial conditions as
the $N$-body runs used a 128$^3$ uniform grid and 8 levels of mesh
refinement, which corresponds to the dynamic range of $128\times
2^8=32768$ and peak resolution of $80/32,768\approx 2.44h^{-1}\ \rm
kpc$. Only the region of $\sim 10h^{-1}\ \rm Mpc$ around the cluster
was adaptively refined, the rest of the volume was followed on the
uniform $128^3$ grid. The dark matter particle mass in the region
around the cluster was $2.7\times 10^{8}h^{-1}{\rm\ M_{\odot}}$, while
other regions were simulated with lower mass resolution.  Note that
although the mass resolution of the gasdynamics and the LR $N$-body
run is similar, the spatial resolution in the latter is higher. In
addition, more aggressive refinement criteria were used in the
$N$-body runs compared to the gasdynamics simulations.  Some
differences in the completeness of subhalo samples are therefore
expected.

The gasdynamics simulations included dynamics of gas and collisionless
DM and several physical processes critical to various aspects of
galaxy formation: star formation, metal enrichment and thermal
feedback due to the supernovae type II and type Ia (SNII/Ia),
self-consistent advection of metals, metallicity- and
density-dependent cooling and UV heating due to cosmological ionizing
background using cooling and heating rates tabulated for the
temperature range $10^2<T<10^9$~K and a grid of densities,
metallicities, and UV intensities using the {\tt Cloudy} code
\citep[ver. 96b4,][]{ferland_etal98}. The cooling and heating rates
take into account Compton heating/cooling of plasma, UV heating,
atomic and molecular cooling.  Star formation in the cluster
simulations is implemented using the observationally-motivated recipe
\citep[e.g.,][]{kennicutt98}: $\dot{\rho}_{\ast}=\rho_{\rm
gas}^{1.5}/t_{\ast}$, with $t_{\ast}=4\times 10^9$~yrs. Stars are
allowed to form in regions with temperature $T<2\times10^4$K and gas
density $n > 0.1\ \rm cm^{-3}$.  A more detailed description of these
simulations will be presented elsewhere.

\section{Halo Identification}
\label{sec:subfind}

A variant of the Bound Density Maxima halo finding algorithm
\citep{klypin_etal99} is used to identify halos and the subhalos
within them.  The details of the algorithm and parameters used in the
halo finder can be found in \citet{kravtsov_etal04}. The main steps of
the algorithm are identification of local density peaks (potential
halo centers) and analysis of the density distribution and velocities
of the surrounding particles to test whether a given peak corresponds
to a gravitationally bound clump. More specifically, we construct
density, circular velocity, and velocity dispersion profiles around
each center and iteratively remove unbound particles using the
procedure outlined in \citet{klypin_etal99}.  We then construct final
profiles using only bound particles and use these profiles to
calculate properties of halos, such as the circular velocity profile
$V_{\rm circ}(r)=\sqrt{GM(<r)/r}$ and compute the maximum circular
velocity $V_{\rm max}$.

If the center of a halo does not lie within a larger system, we
consider the halo to be an isolated or {\it host} halo. In this case
we use the virial radius, $r_{\rm vir}$, defined as the radius
enclosing an overdensity of 180 with respect to the mean density of
the Universe\footnote{We will use this definition of the virial radius
throughout the paper. In \S~\ref{sec:datacomp}, to compare the radial
distributions of galaxies in simulations and observations we will use
$R_{200}$, defined as the radius enclosing the overdensity of 200 with
respect to the {\it critical} density of the Universe.}.  For halos
located within the virial radius of a larger host halo ({\it the
subhalos}), we define the outer boundary at the {\rm truncation
radius}, $r_{\rm t}$, at which the logarithmic slope of the density
profile constructed from the bound particles becomes larger than
$-0.5$ as we do not expect the density profile of the CDM halos to be
flatter than this slope.  Throughout this paper, we will denote the
minimum of the virial mass and mass within $r_{\rm t}$, simply as $M$.
In the gasdynamics simulations, the total mass of an object includes
the mass of dark matter, gas, and stars. For each system we also
estimate the stellar mass, $M_{\ast}$, within the truncation radius.

\begin{figure}[t]
 \vspace{-0.3cm}
 \hspace{3.2cm}
 \centerline{\epsfysize=6.8truein \epsffile{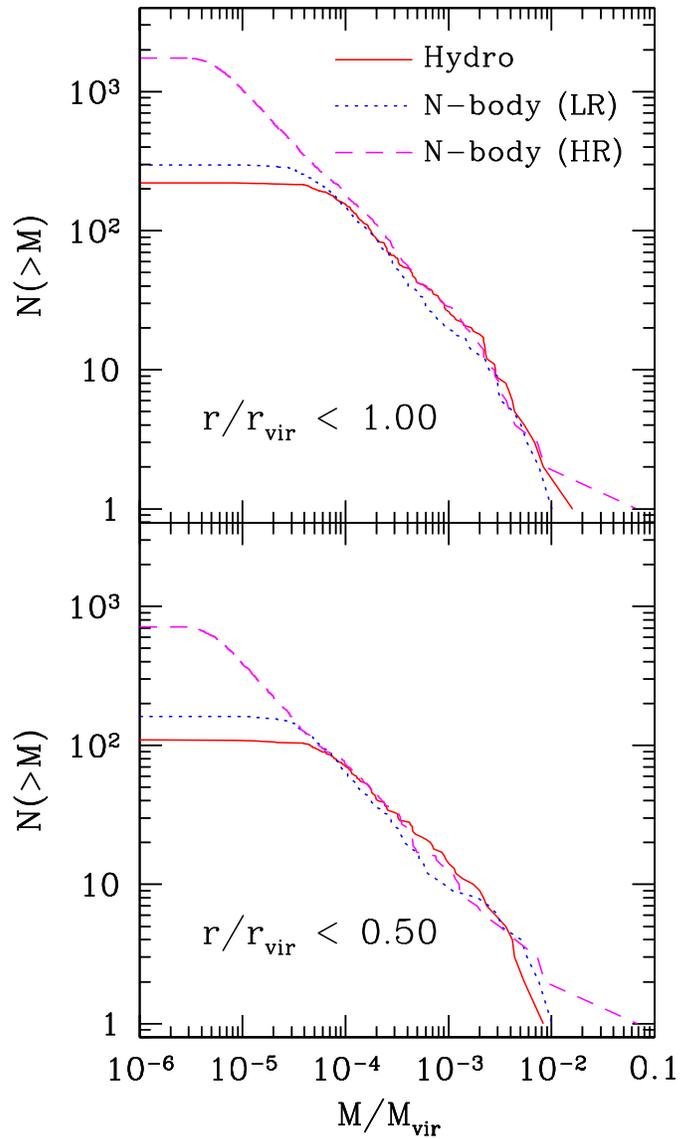}} 
 \vspace{-1.5cm}
\caption{Cumulative mass function of subhalos enclosed within $r_{\rm
    vir}$ ({\it top}) and 0.5$r_{\rm vir}$ ({\it bottom}) of the
  galaxy cluster with $M_{\rm vir}=2.4\times 10^{14}h^{-1}M_{\odot}$.
  Three lines indicate gasdynamics run with cooling and starformation
  ({\it solid}), LR ({\it dotted}), and HR ({\it dashed}) $N$-body
  runs. The mass functions of all three runs converge for $N_{\rm
    p}\gtrsim 80$, which corresponds to $M/M_{\rm vir}>10^{-4}$ for
  hydro and LR runs and $M/M_{\rm vir}\gtrsim 10^{-5}$ for the HR
  $N$-body run. }
\label{fig:mfconv}
\end{figure}

\begin{figure}[t]
\centerline{\epsfysize=3.5truein  \epsffile{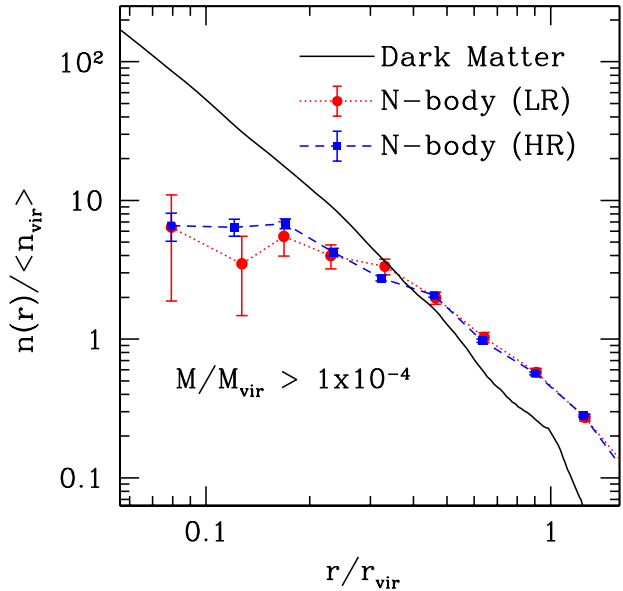}}
\vspace{-0.5cm}
\caption{Convergence test for the radial distribution of subhalos in
  the same cluster as in Figure~\ref{fig:mfconv}. Number density
  profiles of subhalos normalized to the number density within the
  virial radius, $n(r)/\langle n_{\rm vir} \rangle$, in the HR ({\it
    dotted}) and LR ({\it dashed}) $N$-body runs for subhalos with
  masses $M/M_{\rm vir}>10^{-4}$.  The solid line shows the dark
  matter profile normalized to the virial overdensity.  The figure
  shows that the convergence for the radial distribution of subhalos
  is reached, as the profiles are similar in LR and HR runs. }
\label{fig:radialconv}
\end{figure}

\section{Convergence tests}
\label{sec:convtest}

Before we proceed with our analysis, it is important to test the
effects of resolution and determine the smallest mass of objects that
are resolved and identified reliably in our simulations.
Figure~\ref{fig:mfconv} shows comparison of the cumulative mass
functions of subhalos in the two $N$-body and gasdynamics simulations
of the same cluster. We present separate comparisons for the objects
within the virial radius and within the inner half of the virial
radius to show any dependence on radius.

Note that small differences both between gasdynamics and $N$-body
simulations and between low- and high-resolution dissipationless
simulations are expected even at masses where convergence is
reached. The former can arise due to different time integration in the
$N$-body and hydro codes \citep[see,][]{frenk_etal99}, while the
latter can be due to the fact that the HR run includes small-scale
modes absent in the LR run. With this in mind, comparisons of the mass
functions in the LR $N$-body run and gasdynamics run agree well with
the results of the HR $N$-body run for $N_{\rm p}\gtrsim 80$ or
$M/M_{\rm vir}>1\times10^{-4}$. Similar comparison of the velocity
functions of subhalos in the LR $N$-body run agree well with the
results of the HR $N$-body run for $V_{\rm m}/V_{\rm vir}>0.1$ or
$V_{\rm m}>100\rm\ km/s$ for the cluster halo with $V_{\rm
vir}=1065\rm\ km/s$.  We will therefore only consider objects with
masses and maximum circular velocities above these limits in our
analyses.  We find no radial dependence of the slope of the cumulative
mass and velocity functions.  These results are in agreement with the
convergence studies of \citet{diemand_etal04} and \citet{gao_etal04a}.

Figure~\ref{fig:radialconv} shows the radial number density profiles
of subhalos with total masses $>10^{-4}\ M_{\rm vir}$ in the LR and HR
$N$-body runs. The profiles are obtained by averaging over five
(forty) outputs between $z=0.25$ and $z=0$ for the LR (HR) $N$-body
runs, respectively\footnote{The difference in the number of epochs
used in averaging is due to fewer outputs saved in the LR run.}.  For
comparison of dark matter and subhalos, we normalize the profiles to
the mean density within the virial radius, $\langle n_{vir} \rangle$.
The figure shows that the profiles agree at all probed radii. The
convergence is real and the subhalo profiles match without
normalization (in other words, for the same mass threshold $\langle
n_{vir} \rangle$ is the same).  This shows that both the normalization
and shape of the radial profile of subhalos have converged for the
current resolution of the LR N-body run.  We therefore conclude that
the difference between matter and subhalo distributions is real and is
not due to numerical overmerging, in agreement with
\citet{diemand_etal04} and \citet{gao_etal04a}. This is inconsistent
with the semi-analytic models of \citet{taylor_etal03}, which predict
a considerably more radially concentrated distribution of subhalos
than is observed in simulations.

\section{Dissipationless simulations}
\label{sec:Rad_Nbody}

\begin{figure*}[t]
 \vspace{-0.2cm}
 \hspace{-0.4cm}
 \centerline{\epsfysize=6.2truein \epsffile{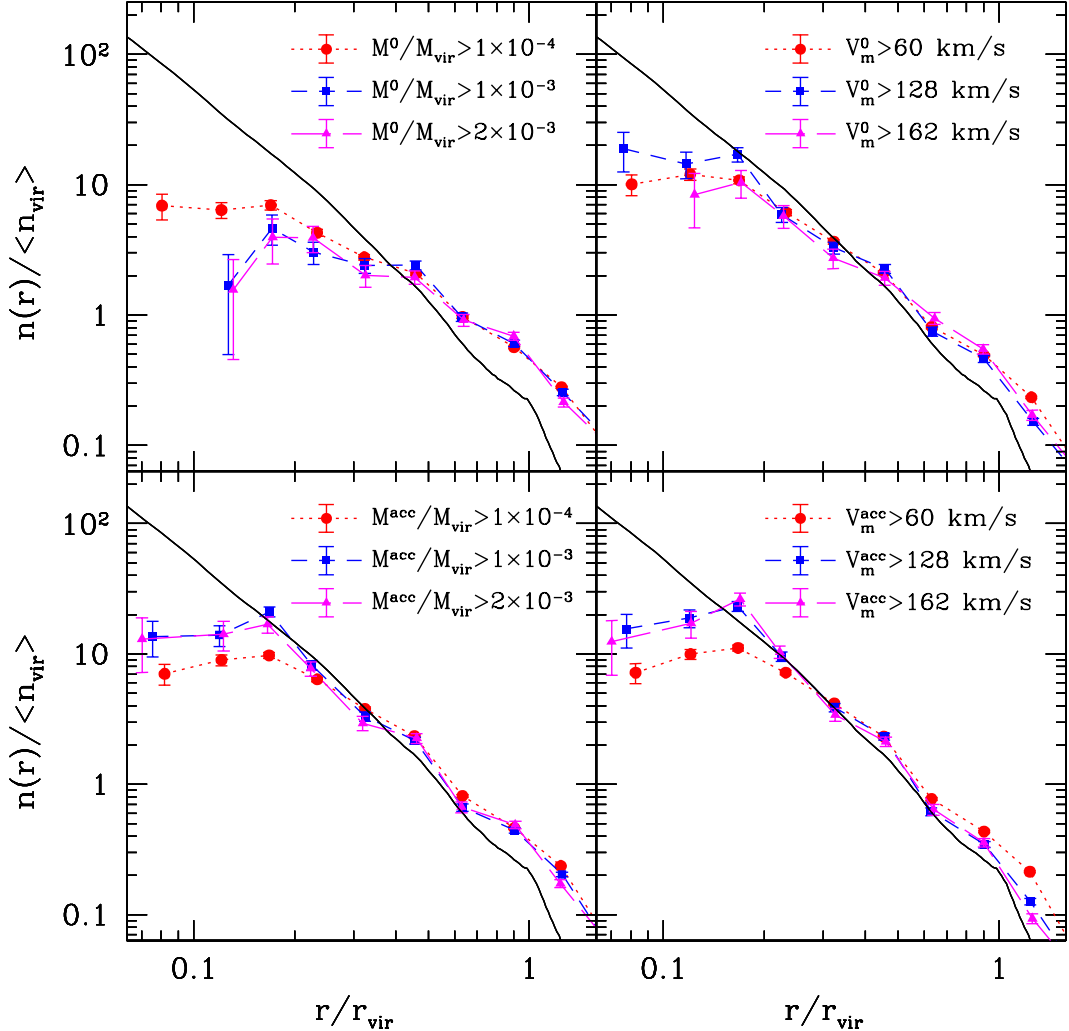}} 
 \vspace{-1.0cm}
\caption{Radial distribution of subhalos for samples with different
  selection criteria.  {\it Top panels:} the subhalos are selected
  using the present-day values of the subhalo total mass, $M^{\rm 0}$,
  (left) and the maximum circular velocity, $V^{\rm 0}_{\rm m}$
  (right).  {\it Bottom panels:} the subhalos are selected using the
  subhalo mass, $M^{\rm acc}$, (left) and the maximum circular
  velocity, $V^{\rm acc}_{\rm m}$ (right) at the time of accretion. In
  each panel the subhalos in the three samples were selected using
  either the minimum mass thresholds of $M/M_{\rm vir}>10^{-4}$ ({\it
  dotted}), $>10^{-3}$ ({\it dashed}) and $>2\times10^{-3}$ ({\it
  long-dashed}) or corresponding values of the circular velocity:
  $V_{\rm m}>60\,\rm km\,s^{-1}$ ({\it dotted}), $>128\,\rm
  km\,s^{-1}$ ({\it dashed}) and $>162\,\rm km\,s^{-1}$ ({\it
  long-dashed}).  The profiles are constructed by stacking 40 outputs
  of the HR $N$-body run between $z=0.25$ and $z=0$.  The error bars
  are the 1$\sigma$ Poisson errors, and the central galaxy is not
  included in the first bin of the profiles. }
\label{fig:radial_Nbody}
\end{figure*}

In this section, we present the analyses of the dissipationless
simulations to study the radial distribution of subhalos and how it
depends on the selection criteria used to define subhalo sample.
Figure~\ref{fig:radial_Nbody} shows the radial number density profiles
normalized to the mean number density of subhalos within the virial
radius in the HR $N$-body run.  The four panels show the radial
distribution of subhalos with different selection criteria. The top
panels show the profile where the subhalos are selected using the
present-day values of the subhalo mass, $M^{\rm 0}$, and the maximum
circular velocity, $V^{\rm 0}_{\rm m}$.  The bottom panels show the
profiles where the values at the time of accretion onto the cluster
are used. These values are obtained from the evolutionary tracks of
each subhalo constructed as described by \citet{kravtsov_etal04b}. We
define the accretion epoch as the time when a halo's distance to the
most massive cluster progenitor first becomes smaller than $1.5$ of
its virial radius (at this epoch).

The profiles are constructed by stacking forty outputs between
$z=0.25$ and $z=0$ to increase the statistics and reduce fluctuations.
Note that this cluster does not evolve significantly between $z=0.25$
and $z=0$. We also checked that the results are not biased by the
stacking procedure.  Throughout the paper, the error bars indicate
1$\sigma$ Poisson error defined by the number of objects in each bin.
Here and throughout, we normalize all profiles to the mean density
within the virial radius because this allows unambiguous comparison of
radial distributions of dark matter and of halos of different masses.
We do not include the central galaxy in the first bin of the profiles.

We show each selection using three values of threshold mass or
circular velocity. The values of the circular velocity thresholds are
chosen to match the corresponding mass thresholds approximately.  The
shape of the profiles does not depend on the subhalo mass and is
significantly shallower than the dark matter distribution for all mass
thresholds.  This is consistent with the results of
\citet{diemand_etal04}.

\begin{figure}[t]
\centerline{ \epsfysize=3.75truein
   \epsffile{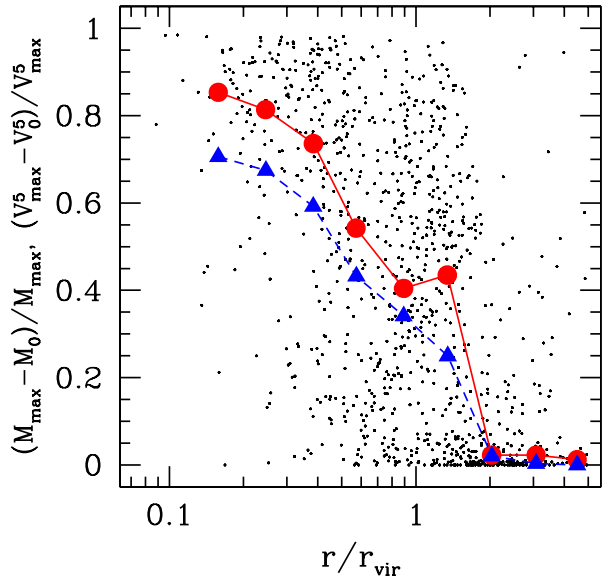} }
\vspace{-0.5cm}
\caption{The fractional mass loss and the change in $V^5_{\rm max}$
  between the epochs when each subhalo has reached the maximum mass,
  $M_{\rm max}$, and $z=0$ as a function of the distance to the
  cluster center.  The {\it dots} show the mass loss experienced by
  individual subhalos, while the {\it solid circles} show the median
  values in logarithmic radial bins.  The {\it triangles} shows the
  same for the change in the fifth power of $V_{\rm max}$. Halos at
  smaller radii have on average experienced considerably more
  stripping than subhalos at larger radii. The tidal mass loss is also
  accompanied by the decrease in $V_{\rm max}$. }
\label{fig:mloss}
\end{figure}

The more extended distribution of subhalos compared to DM in this case
is simply due to the fact that the subhalos in the inner regions have
on average suffered larger tidal mass loss than the halos near the
virial radius.  Figure~\ref{fig:mloss} shows the fractional mass loss
and the change in $V^5_{\rm max}$ experienced by each subhalo since
the epoch when it reached the maximum mass, $M_{\rm max}$, in its
evolution.  The scatter for individual objects is large and
non-Gaussian and is due to the wide distribution of accretion times
and orbital parameters of subhalos.  Nevertheless, there is a clear
average trend of increasing mass loss at smaller radii.  For instance,
the halos within $0.3r_{\rm vir}$ {\it on average} have lost more than
70\% of their mass since accretion. At $r>0.5 r_{\rm vir}$ the halos
on average lose only $\lesssim 40$\% of their mass.  Note also that
the tidal mass loss is also accompanied by a slow decrease in $V_{\rm
max}$ \citep[][see also \citeauthor{tormen_etal04}
\citeyear{tormen_etal04}]{kravtsov_etal04b}.  Mass-based selection
thus biases the subhalo sample to large radii where the tidal
stripping have depleted the number of objects in a given mass range to
a lesser degree.

Figure~\ref{fig:radial_Nbody} shows that the bias introduced by
selecting subhalo samples using circular velocity thresholds, $V_{\rm
  m}^0$, is smaller than in the selection based on mass.  The bias is
expected in this case because circular velocity evolves when a halo
experiences tidal loss as $V_{\rm m}\propto M^{0.2-0.25}$
\citep[][]{kravtsov_etal04b}. The average relation between circular
velocity and mass for isolated halos is $V_{\rm m}\propto M^{0.3}$
\citep[e.g.,][]{avila_reese_etal99,bullock_etal01}. The mass evolution
would thus tend to shift the halos off the relation. Indeed we find
that the normalization of the $M-V_{\rm m}$ relation changes for the
smaller cluster-centric distances, although the slope of the relation
is approximately constant. The slower evolution of the maximum
circular velocity compared to mass means that when it is used to
select subhalos the radial bias is smaller.

If the radial bias is due solely to the varying amount of mass loss at
different radii, we can expect that it should be considerably smaller
if we select subhalos using their mass or circular velocity at the
time of accretion (i.e., not modified by the tidal stripping yet).
The bottom panels of Figure~\ref{fig:radial_Nbody} show the radial
distribution of subhalos for such selections.  It is clear that the
subhalo radial profiles in this case are rather similar to the dark
matter profile at $r\gtrsim 0.2r_{\rm vir}$.  The mass measured at the
accretion, thus, provides a ``label'' for each object, which is not
affected by the subsequent tidal mass loss and evolution.  Note that
we can still expect a certain bias because a larger fraction of
subhalos is fully disrupted by tides at small radii. We cannot
compensate for the absence of the disrupted subhalos with the change
of selection criteria for halos which survived to $z=0$.  We checked
that the results are insensitive to the definition of time of
accretion, as long as subhalos are selected based on their mass and
$V_{\rm m}$ well before entering into the cluster virial region.

Our results therefore indicate that the bias in the radial
distribution of subhalos with respect to dark matter arises simply due
to the radial dependence of the effects of tidal stripping. It is
minimized if we use a property of halos unaffected by the evolution.
In dissipationless simulations such properties can only be computed at
the actual time of accretion if the evolution of a halo is traced in
time.  Unfortunately, the halo properties at the time of accretion are
not observable in reality.  In the next section, we show that the
stellar mass is an observable with similar properties.  The stellar
mass of an object should change little until it is almost fully
destroyed by tides because stars are located in the centers of halos
and are tightly bound. We can expect, therefore, that stellar mass
should correlate well with the mass and circular velocity at the time
of accretion and that the distribution of galaxies selected using
stellar mass (or luminosity) should be similar to that of dark matter.

\vspace{0.75cm}
\section{Gasdynamics simulations}
\label{sec:Rad_Hydro} 

\subsection{Radial distribution of subhalos }

\begin{figure}[t]
  \vspace{-0.5cm}
  \hspace{3.2cm}
  \centerline{\epsfysize=6.8truein \epsffile{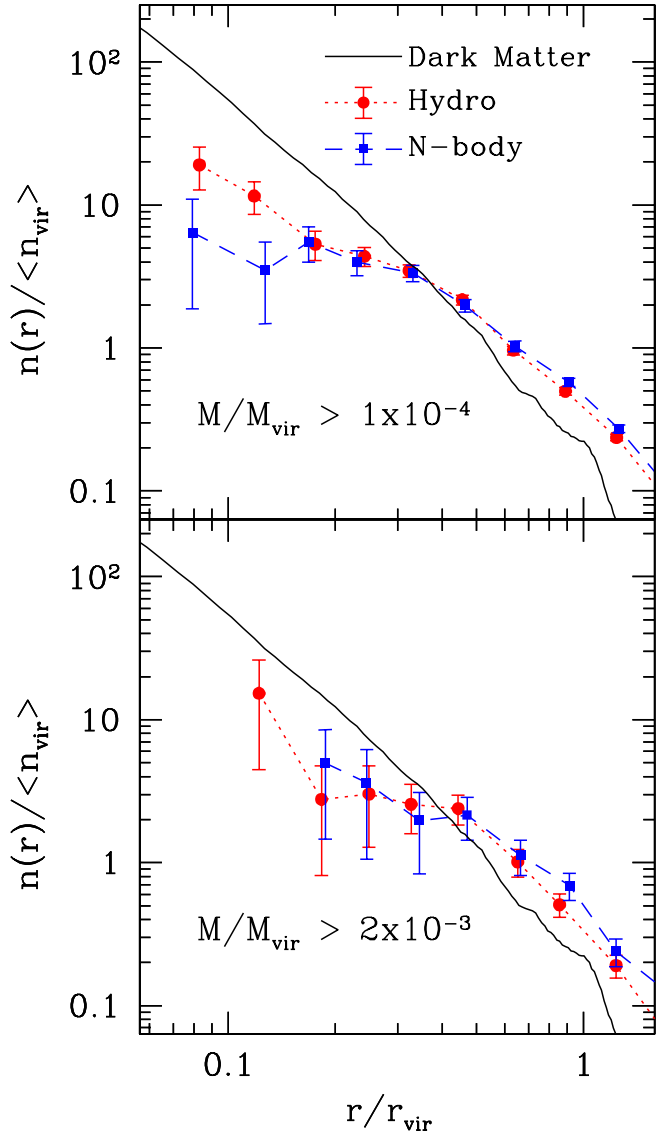}} 
  \vspace{-1.0cm}
\vspace{-0.5cm}
\caption{Radial profiles $n(r)/\langle n_{\rm vir} \rangle$ in
  gasdynamics {\it (dotted)} and $N$-body {\it (dashed)} runs for
  halos with $M/M_{\rm vir}>10^{-4}$ ({\it top}) and $M/M_{\rm
  vir}>2\times10^{-3}$ ({\it bottom}). The {\it solid line} shows the
  dark matter profile in the gasdynamics run. The radial distribution
  of subhalos is remarkably similar for these two runs, regardless of
  the mass cut. This indicates that gas cooling in our simulations
  does not significantly affect the survival of subhalos and their
  radial distribution. }
\label{fig:rad_A}
\end{figure}

In this section we consider the effects of gas cooling and
starformation on the abundance and radial distribution of subhalos.
Figure~\ref{fig:rad_A} compares the radial distribution of subhalos
with $M/M_{\rm vir}>10^{-4}$ and $M/M_{\rm vir}>2\times10^{-3}$ in the
$N$-body and gasdynamics simulations of the same cluster.  These
profiles are constructed by stacking five (nine) outputs between
$z=0.25$ and $z=0$ for $N$-body (gasdynamics) simulation. The radial
distribution of subhalos in the $N$-body and gasdynamics simulations
are remarkably similar.\footnote{Although we plot profiles normalized
to $\langle n_{\rm vir}\rangle$ to compare to the DM profile, the
values of $\langle n_{\rm vir}\rangle$ are the same in the gasdynamics
and dissipationless run, as can be seen in
Figure~\ref{fig:mfconv}. The unnormalized profiles do match in the
same way as in Figure~\ref{fig:rad_A}.} This indicates that the gas
cooling has relatively little effect on the survival and radial
distribution of subhalos. Although cooling can clearly help survival
of subhalos in lower resolution simulations, the halos in our $N$-body
simulations are apparently sufficiently dense to avoid premature
disruption. Given the results of the recent studies of the causes of
the overmerging problem \citep[e.g.,][]{moore_etal96,klypin_etal99},
this is not surprising.  Gas cooling significantly affects mass
distribution only in the inner few percent of the virial radius, while
the survival is largely determined by the halo density within the
radius of $V_{\rm max}$ ($\approx 2r_s$), just outside the affected
regions. We should note that survival in the cluster core is probably
enhanced by cooling.  For the subhalos above the mass threshold of
$M/M_{\rm vir}>10^{-4}$, the radial profile in the gasdynamics
simulation is steeper than in the dissipationless run at $r\lesssim
0.2r_{\rm vir}$.  This, however, does not change the overall
difference between subhalo and DM profiles at larger radii.

\subsection{Radial distribution of galaxies}

\begin{figure}[t]
\centerline{ 
     \epsfysize=3.2truein  \epsffile{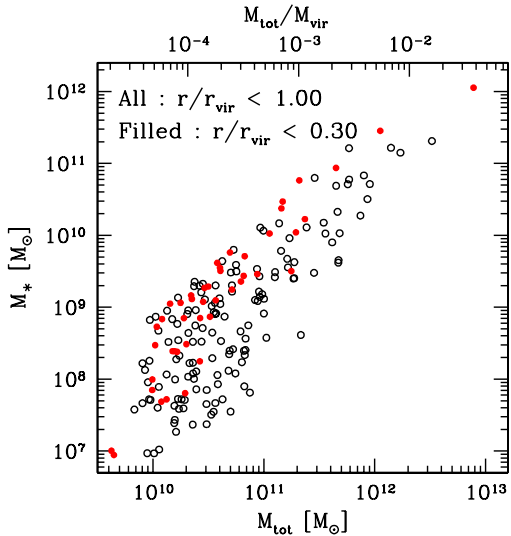}   
}                                                  
\vspace{-0.5cm}
\caption{The mass of stars vs. the total mass of subhalos located within
  $r_{\rm vir}$ (all circles) and $0.3r_{\rm vir}$ (filled circles) of
  the host cluster.  The plot shows that $M_{\ast}$ scales with
  $M_{\rm tot}$ with large scatter, particularly for subhalos with
  $M_{\rm tot}<10^{11}h^{-1}M_{\odot}$. Note that the normalization of
  the relation changes in the inner regions of the cluster, probably due to
  the higher average tidal mass loss of the objects located there. }
\label{fig:masscut}
\end{figure}

\begin{figure}[t]
\centerline{ 
   \epsfysize=3.2truein  \epsffile{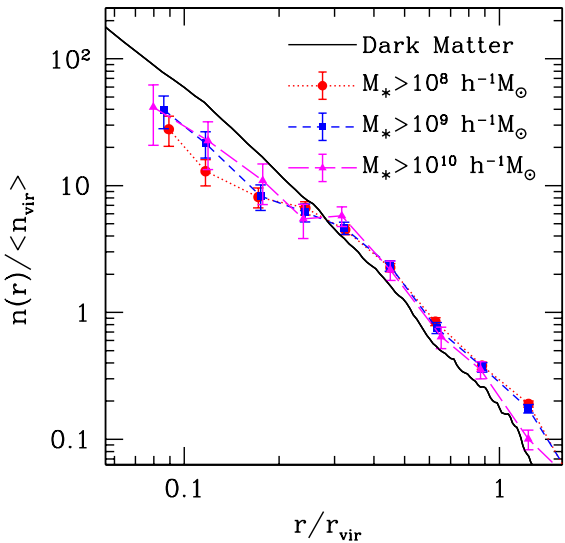}
}
\vspace{-0.5cm}
\caption{ Radial profiles $n(r)/\langle n_{\rm vir} \rangle$ of
  subhalos selected using the stellar mass thresholds of
  $M_{\ast}>10^{8}$ {\it (dotted)}, $10^{9}$ {\it (dashed)} and
  $10^{10}h^{-1}M_{\odot}$ {\it (long-dashed)} in the gasdynamics run.
  Note that the radial distribution of subhalos with the
  $M_{\ast}$-selection does not depend strongly on stellar mass
  threshold and is close to the radial density profile of
  dark matter.}
\label{fig:rad_Hydro}
\end{figure}

\begin{figure}[t]
  \vspace{-0.5cm}
  \hspace{3.2cm}
  \centerline{\epsfysize=6.8truein
    \epsffile{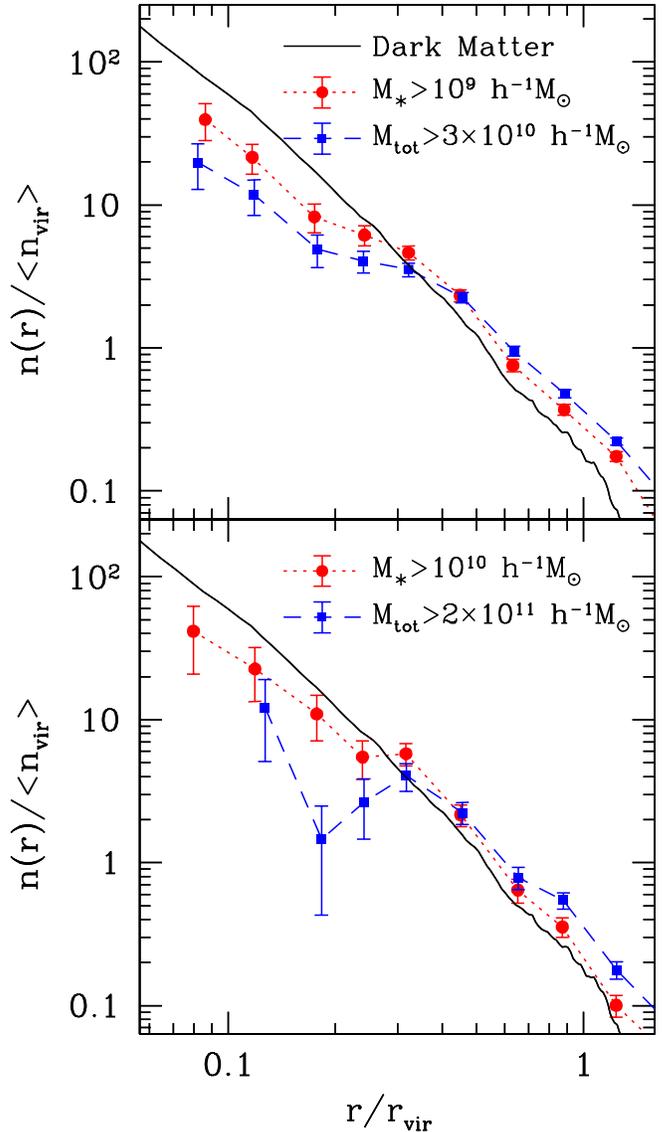}} 
  \vspace{-1.5cm}
\caption{Radial profiles $n(r)/\langle n_{\rm vir} \rangle$ of
  subhalos in the gasdynamics run selected using the stellar mass {\it
  (dotted)} and corresponding total mass {\it (dashed)}. The solid
  line shows the radial density profile of dark matter in the
  gasdynamics simulation. We show the profiles for two different mass
  thresholds: $M_{\ast}>10^9h^{-1}M_{\odot}$ and $M_{\rm
  tot}>3\times10^{10}h^{-1}M_{\odot}$ ({\it top panel}),
  $M_{\ast}>10^{10}h^{-1}M_{\odot}$ and $M_{\rm
  tot}>2\times10^{11}h^{-1}M_{\odot}$ ({\it bottom panel}).  Selection
  using stellar mass results in a significantly steeper radial
  profile. }
\label{fig:masscut2}
\end{figure}

\begin{figure}[t]
  \vspace{-0.5cm}
  \hspace{3.2cm}
  \centerline{\epsfysize=6.8truein
    \epsffile{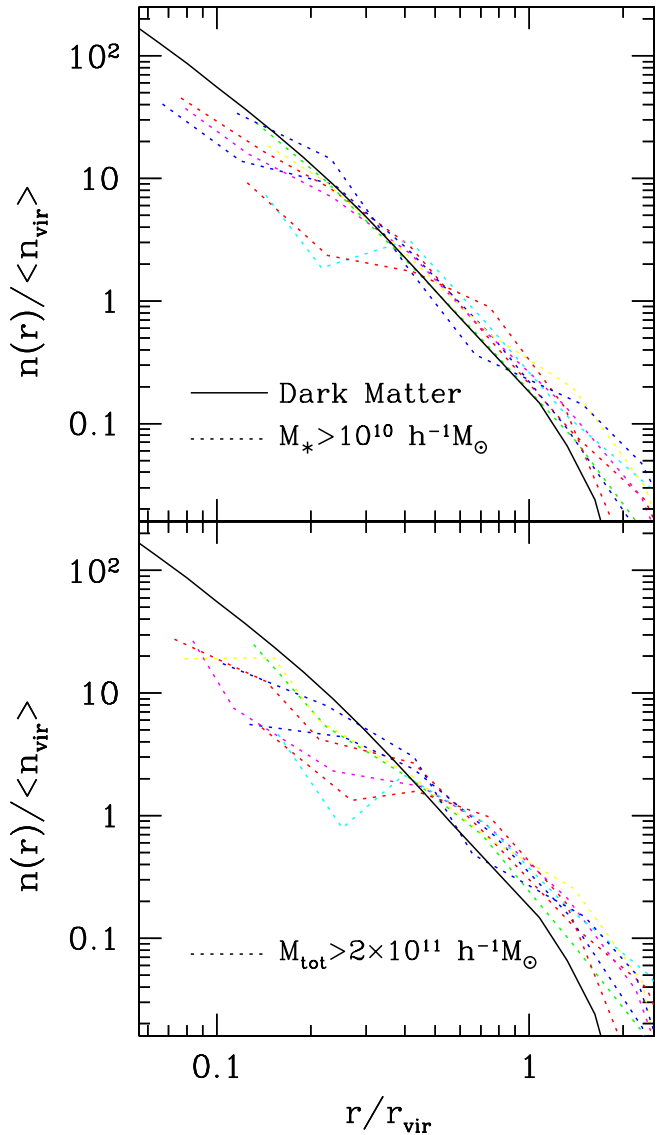}} 
  \vspace{-1.5cm}
\caption{Radial profiles of subhalos $n(r)/\langle n_{\rm vir}
  \rangle$ for the eight simulated clusters at $z=0$.  Here, subhalos
  are selected using their stellar {\it (top)} and corresponding total
  {\it (bottom)} masses of $M_{\ast}>10^{10}h^{-1}M_{\odot}$ and
  $M_{\rm tot}>2\times10^{11}h^{-1}M_{\odot}$ respectively. The solid
  line is the density profile of dark matter averaged over all
  clusters. The figure shows that the distribution of subhalos
  selected using $M_{\ast}$ follows the dark matter profile at
  $0.1<r/r_{\rm vir}<1$ reasonably well. However, there is significant
  variation in the inner slope of the galaxy number density profiles
  among different clusters; for two out of the eight clusters the
  profile in the center is considerably shallower than that of the DM.
}
\label{fig:rad3d_eight}
\end{figure}

In reality, we observe galaxies, not subhalos, and it would be
interesting to study the distribution of galaxies in simulations directly. To
interpret the results, we first need to understand how stellar mass
relates to the total mass of the host subhalo.  We compute the stellar
mass, $M_{\ast}$, as the mass enclosed within the tidal truncation
radius of each subhalo.  Figure~\ref{fig:masscut} shows that the
stellar mass of a galaxy, $M_{\ast}$, correlates with the subhalo
mass, $M_{\ast}\propto M_{\rm tot}^{\alpha}$. The slope of the
correlation ranges in $\alpha\sim 1-1.5$. However, there is
significant scatter, which becomes increasingly larger for the smaller
subhalo masses.  Note that the galaxies in the core of the cluster
($r<0.3r_{\rm vir}$) populate the region of the plot near the upper
envelope of points.  In other words, for a given stellar mass, the
total subhalo mass on average becomes smaller in the inner regions of
cluster.

This systematic trend likely arises because the subhalos
preferentially lose DM mass from their outer regions, while the tidal
mass loss of more tightly bound stars is considerably smaller. In this
case, as the subhalos are stripped and total mass decreases, the
stellar mass stays approximately constant resulting in the evolution
of points horizontally to the left in the $M_{\rm tot}-M_{\ast}$
plane. As we show in Figure~\ref{fig:mloss}, the subhalos in the
cluster core on average have experienced larger tidal mass loss than
the halos near the virial radius, which explains the systematic shift
between solid and open points in Figure~\ref{fig:masscut}. The amount
of tidal mass loss experienced by each object depends on its epoch of
accretion, orbital parameters, and to some extent on its internal
structure. The evolutionary differences among subhalos would explain
the large scatter of the $M_{\rm tot}-M_{\ast}$ correlation.  The
maximum circular velocity of subhalos in simulations with cooling
changes much more slowly than the tidally bound total mass.
Consequently, we find that the correlation of $V_{\rm max}$ and
$M_{\ast}$ is considerably tighter than the $M_{\rm tot}-M_{\ast}$
correlation (Nagai \& Kravtsov 2004, in preparation).

Figure~\ref{fig:rad_Hydro} shows the radial distribution of galaxies
selected using three different stellar mass cuts: $M_{\ast}>10^{8}$,
$10^{9}$, and $10^{10}h^{-1}M_{\odot}$.  The figure shows that the
radial distribution of galaxies with the stellar mass selection does
not depend strongly on stellar mass and is close in shape to the dark
matter profile at $r\gtrsim 0.1-0.2r_{\rm vir}$. Note that in the
simulations with cooling, the DM profile at $r\lesssim 0.1r_{\rm vir}$
is somewhat steeper than the corresponding profile in the
dissipationless simulation \citep{gnedin_etal04}.  It is clear that
$M_{\ast}$-based selection results in a radial distribution of
galaxies which is considerably steeper than the radial profile of
subhalo samples selected using total mass.

Figure~\ref{fig:masscut2} shows this more clearly using direct
comparison of the radial distribution of objects selected using
stellar and corresponding total mass.  The thresholds in $M_{\rm tot}$
are chosen to correspond approximately to the $M_{\ast}$ thresholds
using the $M_{\rm tot}-M_{\ast}$ correlation
(Figure~\ref{fig:masscut}).  The figure shows that the radial
distribution of objects in $M_{\ast}$-selected samples is
significantly steeper than the profiles of objects selected by the
total mass. As expected, the difference is similar to the difference
of profiles of subhalos selected using total mass at $z=0$ and at the
time of accretion, seen in \S~\ref{sec:Rad_Nbody}.  The stars are
located near the centers of DM subhalos and are tightly bound. The
stellar mass thus should not evolve significantly even if an object
sheds a large fraction of its DM halo \citep[e.g.,][and references
therein]{gnedin03a}.  Our results show that this is indeed the case.

\begin{figure}[t]
  \centerline{\epsfysize=3.2truein \epsffile{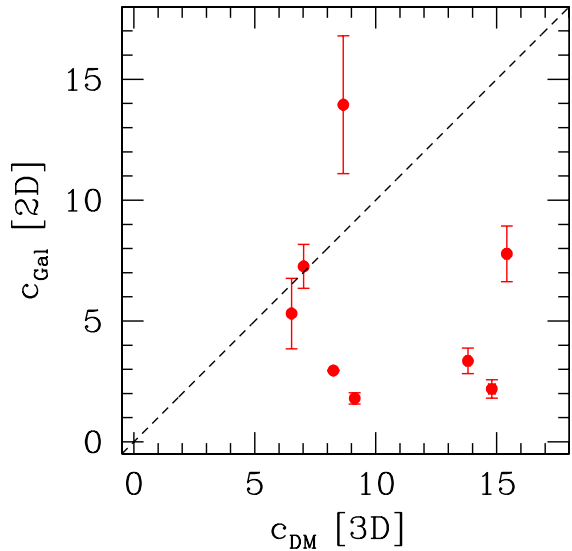}}
\caption{Comparison of the concentration parameters of galaxies,
$c_{\rm Gal}$, and dark matter, $c_{\rm DM}$, for the eight simulated
clusters at $z=0$.  $c_{\rm Gal}\mbox{[2D]}$, is obtained by fitting the
NFW profile to the projected (2D) galaxy distribution, whereas $c_{\rm
DM}\mbox{[3D]}$, is the concentration parameter of dark matter
obtained by fitting the 3D dark matter profile.}
\label{fig:con}
\end{figure}

Since the results so far are based on the analyses of one cluster, we
investigate if there is a variation in the distribution of galaxies
among different clusters.  Figure~\ref{fig:rad3d_eight} shows the 3D
radial distribution of galaxies for the sample of eight clusters at
$z=0$.  We find that for all clusters the radial distribution of
galaxies selected based on the stellar mass, $M_{\ast}$, is steeper
than that of the samples selected using the total mass, $M_{\rm tot}$.
The figure also illustrates significant variation in the inner slope
of the profile among different clusters. The radial profiles of
galaxies in most clusters are close to the dark matter profiles at all
radii, but there are some that show significant flattening of the
galaxy number density profile at $r\lesssim 0.3r_{\rm vir}$.  We find
that the variation among profiles does not decrease if we use clusters
at the epochs when they are most relaxed.  This indicates that the
flattening of the galaxy number density profile is not related to the
host halo mass or to the dynamical state of clusters but may be
related to their evolutionary histories.

It would also be interesting to ask whether the concentration of the
radial profile of galaxies correlates with the concentration of the DM
distribution. In Figure~\ref{fig:con} we compare the concentration
parameters of galaxies, $c_{\rm Gal}$, and dark matter, $c_{\rm DM}$,
for the eight simulated clusters at $z=0$.  For galaxies we use the
concentrations obtained by fitting the projected Navarro-Frenk-White
profile \citep[hereafter NFW,][]{navarro_etal97} to the projected (2D)
galaxy distribution, because the projected profiles are directly
observable.  For the dark matter, on the other hand, we use the true
concentration obtained by fitting the NFW to the 3D dark matter
distribution.  The details of the fitting procedure are described in
the Appendix.

The concentration parameters exhibit significant variation among
different clusters.  The concentration parameters of dark matter span
a wide range from 6 to 16.  Note that the concentration of dark matter
in the current simulations may be systematically larger than the
prediction of the dissipationless simulations
\citep{navarro_etal97,bullock_etal01,eke_etal01,tasitsiomi_etal04a}
because the baryon dissipation steepens the inner profiles of dark
matter \citep{gnedin_etal04}.  Two of the eight clusters have very
similar best fit concentrations of galaxy and DM profiles. There is
also one cluster for which the concentration of galaxy profile is
larger than that of dark matter. Nevertheless, the concentration
parameter of galaxies in the majority of clusters is smaller than that
of dark matter. These results indicate that fits to the projected
radial profiles of galaxies may be a poor probe of the DM
concentrations.

\subsection{Comparison to observations}
\label{sec:datacomp}

\begin{figure}[t]
  \centerline{\epsfysize=3.2truein \epsffile{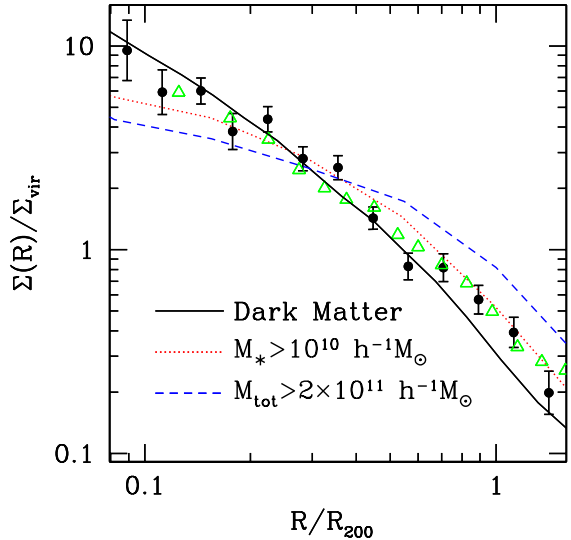}}
  \vspace{-0.5cm}
\caption{Projected radial distribution of galactic subhalos
  $\Sigma(R)/\Sigma_{vir}$ averaged over the eight simulated clusters
  at $z=0$ and three orthogonal projections.  We show the radial
  distribution of subhalos selected using the mass thresholds of
  $M_{\ast}>10^{10}h^{-1}M_{\odot}$ ({\it dotted}) and $M_{\rm
    tot}>2\times10^{11}h^{-1}M_{\odot}$ ({\it dashed}).  The solid
  line is the average projected profile of dark matter in the
  gasdynamics simulation.  The {\it solid circles} are the average
  radial profile of galaxies in clusters in the CNOC survey
  \citep{carlberg_etal97}, and the {\it open triangles} are the
  average profile in the 2MASS cluster survey \citep{lin_etal04}.  The
  data points are scaled arbitrarily.  Note that the distribution of
  $M_{\ast}$-selected subhalos is consistent with the observed
  distribution of galaxies over the entire range of probed radii:
  $0.1<R/R_{\rm 200}<2.0$.}
\label{fig:rad2d_comp}
\end{figure}

Here we compare the distribution of galaxies and subhalos in the
$\Lambda$CDM clusters in our simulated sample to the observed galaxy
distribution.  Figure~\ref{fig:rad2d_comp} shows the projected radial
distribution of dark matter, galaxies and subhalos averaged over the
eight simulated clusters at $z=0$.  The profiles are constructed by
projecting spherical volume with radius of 3$r_{\rm vir}$ centered on
the cluster center and averaging over all simulated clusters and three
orthogonal projections of each cluster.  The profiles are normalized
in units of the mean surface density of dark matter, $\Sigma_{\rm
vir}=M_{\rm vir}/\pi r_{\rm vir}^2$, and galaxies/subhalos, $N_{\rm
vir}/\pi r_{\rm vir}^2$, where $M_{\rm vir}$ and $N_{\rm vir}$ are the
mass and the number of objects enclosed within the sphere with the
virial radius $r_{\rm vir}$ respectively.  The data points are from
the CNOC \citep{carlberg_etal97} and 2MASS \citep{lin_etal04} cluster
surveys.  The observed galaxy profiles are scaled arbitrarily and the
2MASS survey profile has been re-binned to smaller number of bins for
clarity.

The figure shows that the projected distribution of galaxies in both
simulations and observations is more concentrated than that of
subhalos selected using the total mass. The radial profile of galaxies
is somewhat shallower than the average dark matter distribution.
Nevertheless, the figure shows that the distribution of galaxies in
the simulation is in reasonably good agreement with the observed
distribution of galaxies for the entire probed range of scales:
$0.1<R/R_{\rm 200}<2.0$.  The agreement with the data is considerably
better if two clusters with very flat inner profiles (see
Figure~\ref{fig:rad3d_eight}) are excluded.

\section{Discussion and conclusions}
\label{sec:discussion}

We presented a study of the radial distribution of subhalos and
galaxies using high-resolution cosmological simulations of galaxy
clusters formed in the concordance $\Lambda$CDM cosmology. In this
paper, we first analyze the dissipationless simulations to study
properties and processes that govern the evolution of subhalos. We
then analyze cluster simulations with gasdynamics, cooling, and
starformation started from the same initial conditions to study the
effect of these processes on the spatial distribution of subhalos and
galaxies.

In agreement with previous studies, we find that the radial profile of
subhalos within the virial radius in the dissipationless simulations
is significantly shallower than the dark matter distribution, if the
subhalos are selected based on the present-day values of the subhalo
mass, $M^{\rm 0}$.  Comparing two simulations of the same cluster with
an order of magnitude different resolution, we show that both the
radial profile of subhalos and DM profiles have converged at the
scales analyzed here.  The difference between matter and subhalo
distributions is therefore real and is not due to numerical
overmerging, in agreement with conclusions of \citet{diemand_etal04}.
Selection on the present-day value of the maximum circular velocity,
$V^{\rm 0}_{\rm m}$, results in steeper radial profile of subhalos.
The radial bias with respect to the dark matter distribution in this
case is weaker than in the case of mass-based selection.

We show that the radial bias of subhalo distribution is due to the
tidal mass loss experienced by subhalos as they orbit in the cluster
potential. We find that subhalos in the inner regions have on average
suffered larger tidal mass loss than the halos near the virial radius
\citep[see also][]{gao_etal04a}.  For instance, the halos within
$0.3r_{\rm vir}$ {\it on average} have lost more than 70\% of their
mass since accretion, while subhalos at $r>0.5 r_{\rm vir}$ on average
lose only $\lesssim 40$\% of their mass.  Mass-based selection,
therefore, biases the subhalo sample to large radii where tidal
stripping have depleted the number of objects in a given mass range to
a lesser degree.  The mass loss is also accompanied by decrease in the
maximum circular velocity, but the decrease is slower compared to
mass.  The slower evolution of the maximum circular velocity compared
to mass means that when it is used to select subhalos the radial bias
is smaller.

One of the obvious omissions in dark matter only simulations is the
lack of dissipation.  One can thus ask whether the baryon dissipation
makes the subhalo more resistant to tidal disruption.  If so, is the
effect strong enough to resolve the differences between matter and
subhalo distribution?  Comparing the simulations with and without
cooling and starformation, we find that the baryon dissipation indeed
increases the survival of subhalos, especially in the inner regions of
clusters.  However, the effect is relatively small because the baryon
dissipation makes the core more resistant to the tidal disruption, but
not the entire structure of the subhalo.  The efficiency of tidal
losses for the total mass is, therefore, largely unaffected by the
baryon dissipation except in the center of the subhalos.  Thus, we
conclude that this effect alone does not resolve the differences
between matter and subhalo distributions.

If the radial bias is due solely to the varying amount of mass loss at
different radii, the bias can be minimized if we use a property of
halos unaffected by the evolution. In dissipationless simulations such
property can be computed at the actual time of accretion if the
evolution of a halo is traced in time. For instance, the radial
profiles of subhalos selected using mass or maximum circular velocity
measured at the time of subhalo accretion onto cluster are very close
to the cluster DM profile. The radial distribution of subhalos and
inferences from comparisons with observed radial distribution of
galaxies in clusters therefore depend on the selection criteria used
to define subhalo sample. Consequently, the differences between
subhalo and galaxy radial distributions do not necessarily indicate
that dissipationless simulations significantly underestimate the
number of galactic halos in clusters
\citep[e.g.,][]{diemand_etal04,gao_etal04b}.  They may simply indicate
a difference in selection criteria for subhalo and galaxy samples.

Unfortunately, halo properties at the time of accretion are not
observable in reality. The stellar mass, however, is an observable
that can behave similarly.  Stellar mass of a galaxy should change
little until it is almost fully destroyed by tides because stars are
located in the centers of halos and are tightly bound. The total mass,
on the hand, can decrease dramatically due to tidal stripping.  Using
the simulations that include cooling and starformation, we show that
the radial distribution of subhalos selected based on the stellar mass
is considerably steeper than the that of subhalos selected using the
total masses in the same simulation.

We show that the profiles of galaxies in the simulations are in good
agreement with the observed projected distribution of galaxies for the
entire radial range probed by the current simulations: $0.1<R/R_{\rm
  200}<2.0$.  The NFW fits to the average profiles of eight simulated
clusters at $z=0$ in the range $0.1<R/R_{\rm 200}<1.0$ give the
best-fit concentration parameter of $c_{\rm Gal}\approx 2-3$ for
galaxies and $c_{\rm DM}\approx 10$ for the dark matter. In other
words, the radial profile of galaxies selected using the stellar mass
is, on average, somewhat more extended than the dark matter profile.
Note that the values of $c_{\rm Gal}$ in our simulations are
consistent with the observational estimates of $c_{\rm Gal}\sim 3-4$
\citep{carlberg_etal97,vandermarel_etal00,lin_etal04}. There are 
indications that in group-size systems galaxies have more extended
radial profiles than DM \citep{vandenbosch_etal04,mathews_etal04}. The
concentration of DM profiles are expected to be in the range $c_{\rm
  DM}\sim 5-10$
\citep{navarro_etal97,bullock_etal01,eke_etal01,tasitsiomi_etal04a} or
larger if the profiles are significantly affected by gas cooling
\citep{gnedin_etal04}. We do not find any strong correlation between
galaxy and DM concentrations in simulated clusters (see
Figure~\ref{fig:con}). Thus, our results imply that the projected radial
profiles of galaxies in clusters may, in general, be a poor probe of
the underlying dark matter distribution.

The fact that there is a difference between galaxy and DM radial
profiles indicates that galaxies are affected by tides in the dense
environment of the cluster cores. Higher-resolution simulations will
be needed to test whether the amount of the stellar mass loss and
efficiency of tidal disruption have converged.  If the flattening of
the inner profile of galaxies is due to real tidal mass loss or
disruption experienced by galaxies, we expect to find a large amount
of stellar debris in the cluster core.  Therefore, the detailed
studies of the properties and origin of the stellar debris and
intracluster light should provide new insights into the connection
between the evolution of galaxies in clusters and the formation of cD
galaxies and their stellar envelopes.

\acknowledgments

We would like to thank Stefan Gottl\"ober for giving us the HR
re-simulation of one of the clusters, and Argyro Tasitsiomi for
providing the NFW fitting code. AVK would like to thank Aspen Center
for Physics and organizers of the ``Starformation in galaxies''
workshop for hospitality and productive atmosphere during completion
of this paper. This project was supported by the National Science
Foundation (NSF) under grants No.  AST-0206216 and AST-0239759, by
NASA through grant NAG5-13274, and by the Kavli Institute for
Cosmological Physics at the University of Chicago. D.N. is supported
by the NASA Graduate Student Researchers Program and by NASA LTSA
grant NAG5--7986.  The cosmological simulations used in this study
were performed on the IBM RS/6000 SP4 system at the National Center
for Supercomputing Applications (NCSA) and at the Leibniz
Rechenzentrum Munich and the John von Neumann Institute for Computing
J\"ulich.

\bibliography{rad}

\appendix
\section{NFW profile fitting procedure}
\label{sec:fits}
 
For each halo, we fit the NFW analytic density
profiles of the form
\begin{eqnarray}
\label{eq:nfw1}
\rho(r) & = & \frac{\rho_s}{x (1+x)^{2}},   \\ \nonumber
\Sigma(r) & = & \frac{2 \rho_s r_s}{x^2 -1} f(x),
\end{eqnarray}
where with $x\equiv r/r_{s}$.  The former is the 3D analytic NFW
profile.  The latter is the analytic 2D profile obtained by projecting
the 3D profile along the line-of-sight from the negative to positive
infinity, and $f(x)$ is given by \citep{bartelmann96}:
\begin{equation}
\label{eq:nfw2D_g}
f(x) = \cases{
  1 - {2\over\sqrt{x^2 - 1}}{\rm arctan}\sqrt{x-1\over x+1} & $(x>1)$ \cr
  1 - {2\over\sqrt{1 - x^2}}{\rm arctanh}\sqrt{1-x\over 1+x} & $(x<1)$ \cr
  0 & $(x=1).$ \cr
  }
\end{equation}

The NFW profile has two free parameters $r_s$ and $\rho_s$ or,
equivalently, the concentration parameter $c_{\rm vir}\equiv r_{\rm
vir}/r_{\rm s}$ and the virial radius $r_{\rm vir}$ of the halo.
These parameters are highly degenerate, which makes the fits of the
analytic profiles to the profiles of the simulated clusters sensitive
to a number of factors, including the choice of binning, the merit
function, the range of radii used in the fitting, the weights assigned
to the data points. As we are only interested in the concentration
parameters of galaxies and dark matter of the simulated clusters, we
can improve the quality of the fits by removing the degeneracy between
$r_{\rm vir}$ and $c$.  This is possible because we can measure
$r_{\rm vir}$ directly in the simulations, leaving concentration as
the only one free parameter.

\begin{figure}[t]
  \centerline{\epsfysize=3.2truein \epsffile{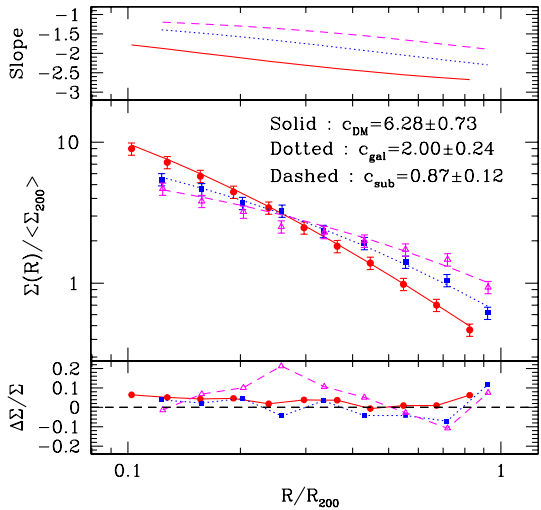}}
  \vspace{-0.5cm}
\caption{{\it Middle panel:} Results of the NFW fits to the projected
  radial distribution $\Sigma(R)/\Sigma_{vir}$ of dark matter and
  galaxies averaged over the eight simulated clusters at $z=0$ and
  three orthogonal projections.  The fits were done using the range
  $[0.1r_{\rm vir},r_{\rm vir}]$ and $\chi^2$ merit function.  Three
  lines indicate dark matter ({\it solid}) and galaxies selected using
  the thresholds of $M_{\ast}>10^{10}h^{-1}M_{\odot}$ ({\it dotted})
  and $M_{\rm tot}>2\times10^{11}h^{-1}M_{\odot}$ ({\it dashed}).
  {\it Bottom panel:} fractional deviations of each best fit NFW
  profile from the actual profile. {\it Top panel:} local logarithmic
  slope as a function of radius for the three fits.  The distribution
  of both dark matter and galaxies is well-described by the NFW
  profile in the radial range $0.1<R/R_{200}<1.0$.  We find that the
  radial distribution of galaxies is shallower than that of dark
  matter, with the concentration parameter of 6.3 for dark matter and
  concentration of 2.0 and 0.87 for subhalos selected using $M_{\ast}$ and
  $M_{\rm tot}$, respectively.}
\label{fig:rad2d_fit}
\end{figure}

When fitting the profiles of the simulated clusters using the
one-parameter NFW analytic profile, we first normalize the radius 
to the virial radius, $y \equiv r/r_{\rm vir}$.  For the 3D
profile, we then need to normalize the analytic profiles in units of
the mean mass density within the virial radius, $\rho_{\rm vir}=3
M_{\rm vir}/ 4 \pi r_{\rm vir}^3$.  Similarly, for the projected (2D)
profile, we normalize the profile in units of the projected surface
mass density enclosed within the sphere with the virial radius,
$\Sigma_{\rm vir}=M_{\rm vir}/\pi r_{\rm vir}^2$.  With this
normalization, the NFW analytic density profiles in Eq.~(\ref{eq:nfw1})
become
\begin{eqnarray}
\label{eq:nfw2}
\frac{\rho(r)}{\rho_{\rm vir}} & = & \frac{1}{3} \frac{c^3 g(c)^{-1}}{cy (1+cy)^{2}}, \\ \nonumber
\frac{\Sigma(r)}{\Sigma_{\rm vir}} & = &
\frac{1}{2} \frac{c^2 g(c)^{-1}}{(cy)^2-1} f(cy),
\end{eqnarray}
with only one free parameter, $c$. Here, $g(c)={\rm ln}(1+c)-c/(1+c)$.

For each halo, we obtain the best fit concentration parameter $c$ by
fitting the analytic profiles given by Eqs.~(\ref{eq:nfw2}) to the 3D
and 2D radial profiles of dark matter, galaxies and subhalos in the
simulations by the standard $\chi^2$ minimization. We use equal-size
logarithmic bins and weigh each bin by the Poisson noise in the number
of dark matter particles, galaxies or subhalos \citep[see][for more
details]{tasitsiomi_etal04a}.  For both dark matter and
galaxies/subhalos, we perform fits using the 12 equal-size logarithmic
bins between 0.1$<r/r_{\rm vir}<$1.0. The fitting region is chosen to
maximize the statistics of galaxies, since the radial profiles of
galaxies are much noisier than the dark matter profile.  The lower
radial bound is set by the fact that the halo finder fails to find
galaxies within $r/r_{\rm vir}\lesssim 0.1$, while the outer bound is
the virial radius of the cluster, beyond which the presence of strong
fluctuations due to cluster substructure significantly biases fits of
smooth analytic profiles.  We checked that the results are not
sensitive to the variation around the chosen number of bins and radial
range.  For example, decreasing the lower/upper bound by a factor of
two change the resulting best fit concentration of dark matter by only
about 10\%. Note that the insensitivity of the results to the
perturbation around the chosen fit parameters is partly because we
removed the degeneracy between $c$ and $r_{\rm vir}$ by measuring
$r_{\rm vir}$ directly from the simulations.  In fact, the same
changes affect the results of the two-parameters fits much more
dramatically.  Figure~\ref{fig:rad2d_fit} shows the best-fit
concentration parameters of the dark matter and galaxy distribution
averaged over eight simulated clusters at $z=0$.  We find that the
concentration parameter is 6.3 for dark matter, 2.0 for galaxies
(selected using $M_{\ast}$) and 0.87 for subhalos (selected using
$M_{\rm tot}$). Note that the average concentration parameter of DM
obtained here is lower than the average value of $\approx 10$ quoted
in \S~\ref{sec:discussion}. This is because fitting of the projected
profiles generally biases the profile due to the projection from the
large distances outside the virial radii, as we did not attempt to
subtract the background in these two dimensional fits.

\end{document}